\documentclass[10pt,a4paper]{book}
\usepackage{udineHyderabad}
\begin{document}
\talktitle{Cosmological  Theories of Special and General Relativity - I}

\talkauthors{Moshe Carmeli \structure{a}}
\authorstucture[a]{Department of Physics, 
                   Ben Gurion University  of the Negev, 
                   Beer Sheva 84105, Israel}

\shorttitle{Theories of Special and General Relativity - I} 

\firstauthor{M. Carmeli}

\begin{abstract}
In the standard cosmological  theory one uses the Einstein  concepts of space 
and time as were originally introduced for the special theory of relativity
and the general relativity theory. According to this approach all
physical quantities are described in terms of the continuum spatial 
coordinates and time. Using general relativity theory a great progress
has been made in understanding the evolution of the Universe. Cosmologists 
usually measure spatial distances and redshitfs of faraway galaxies as 
expressed by the Hubble expansion. In recent years this fact was undertaken
to develop new theories in terms of distances and velocities (redshift). 
While in Einstein's relativity the propagation of light plays the 
major role, in the new theory it is the expansion of the Universe that takes 
that role and appears at the outset. The cosmic time becomes crucial in these
recent theories, which in the standard theory is considered to be
absolute but here it is relative. In this lecture this new approach to cosmology is presented.   
\end{abstract}

\section{Introduction}
It is well known that both Einstein's theories are based on the fact that light
propagates at a constant velocity. However, the Universe also expands at a
constant rate when gravity is negligible. Moreover, cosmologists usually
measure spatial distances and redshifts of 
faraway galaxies as expressed by the Hubble expansion. In recent years this
fact was undertaken to develop new theories in terms of distances and velocities
(redshift). While in Einstein's special relativity the propagation of light  
plays the major role, in the new theory it is the expansion of the Universe 
that takes that role. It is the concept of cosmic time that becomes crucial in 
these recent theories. In the standard theory the cosmic time is considered to 
be absolute. Thus we talk about the Big Bang time with respect to us here on
Earth as an absolute quantity. Consider, for example, another galaxy that 
has,
let us say, a relative cosmic time with respect to us of 1 billion year. Now 
one may ask what will be the Big Bang time with respect to this galaxy. Will
it be the Big Bang time with respect to us minus 1 billion year? A person
who lives in that galaxy will look at our galaxy and say that ours is far away 
from him by also 1 billion year. Will that mean, with respect to him, our
galaxy is closer to the Big Bang time by 1 billion year? Or will we seem to 
him to be farther by 1 billion year? All this leads to the conclusion that 
there is no absolute cosmic time. Rather, it is a relative concept.

Based on this assumption, we present in this lecture a theory that relates
distances between galaxies and their relative velocities. These are actually 
the dynamical variables that astronomers measure. In the first part the theory
will be relating distances to velocities with very  weak gravitational field
(special relativity). We then, in the second part, extend the theory to 
include
the gravitational field of the Universe. Before doing that, we present a brief
review of Einstein's special relativity theory.
\section{Einstein's special relativity: A review}
We will not give a full detail of this very important theory. Rather, we 
outline some of its fundamentals [1,2]
1. The Michelson-Morley experiment for the constancy of the speed of light.
\newline
2. Einstein's two postulates: (a) Constancy of the speed of light; (b) 
Validity of the laws of physics in internal coordinate systems. (See Figure 1)
\begin{figure}[t]
\centering
\includegraphics[height=10cm,angle=180]{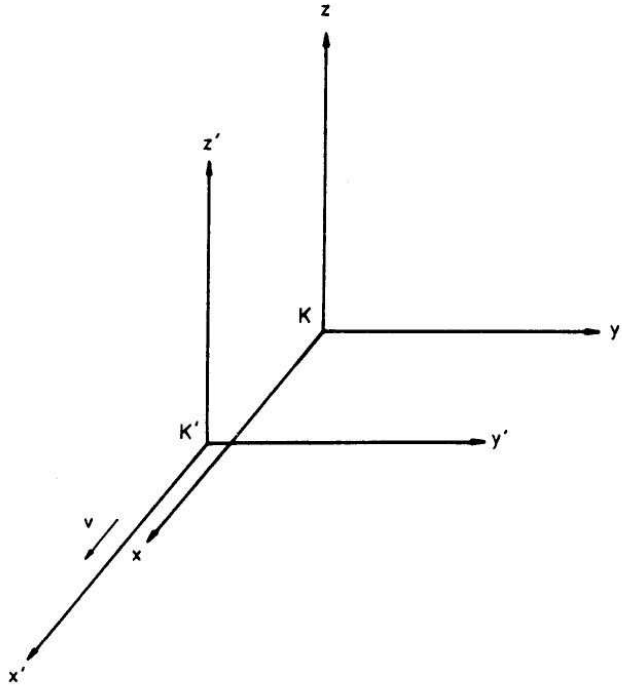}
\caption{Two coordinate systems $K$ and $K'$, one moving with respect to the 
other with a velocity $v$ in the $x$-direction.}
\label{fig.1}
\end{figure}
\newline
3. The Lorentz transformation. This is given by 
$$ct'=\left(ct-\beta x\right)/\sqrt{1-\beta^2},\hspace{5mm}
x'=\left(x-\beta ct\right)/\sqrt{1-\beta^2},\eqno(2.1)$$
where $\beta=v/c$, and $y'=y$, $z'=z$.\newline
4. Minkowski's unification of space and time.\newline
5. Invariance of the laws of physics under the Lorentz transformation.\newline
6. Minkowski's line element
$$ds^2=c^2dt^2-\left(dx^2+dy^2+dz^2\right)\eqno(2.2)$$
7. The light cone (see Figure 2).
\begin{figure}[t]
\centering
\includegraphics[height=10cm]{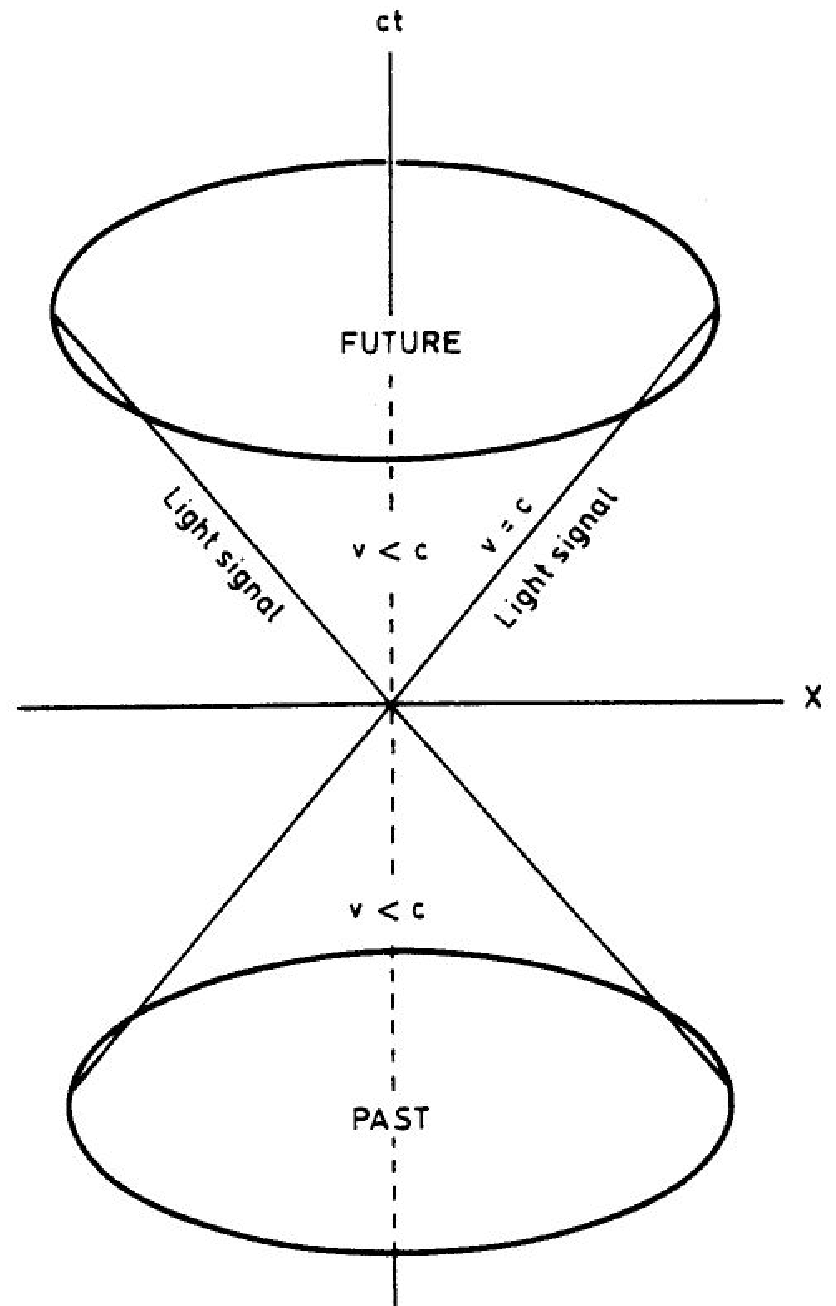}
\caption{The light cone in two dimensions, $x^0(=ct)$ and $x^1(=x)$. The 
propagation of two light signals in opposite directions passing through 
$x=0$ at time $t=0$, is represented by the two diagonal straight lines. 
(Compare the galaxy cone given in the sequel.)}
\label{fig.2}
\end{figure}
\section{Cosmological special relativity}
We outline this theory very briefly with the following 
points \cite{Carm95,Carm96,Carm02}.\newline
1. Hubble's law $R=\tau v,$
where $R$ is the distance to a galaxy, $v$ is the receding velocity of the 
galaxy,
$\tau$ is universal constant equal to 12.486 Gyr (Big Bang time).\newline
2. Cosmic time is not absolute but a relative concept.\newline
{\it Example: Another galaxy is one billion year with respect to us. One may 
ask, what is the Big Bang time with respect to this galaxy. Will it be BB time
with respect to us minus 1 billion year?From the point of view of that galaxy 
ours is far away from it by also 1 billion year. Does that mean, with respect 
to that galaxy, ours is closer to the BB time by 1 billion year?Or our galaxy
will seem to be farther by 1 billion year? All this leads to the conclusion 
that there is no absolute cosmic time. Rather, it is a relative concept.}\newline
3. Universe with negligible gravity.\newline
4. Line element of an expanding Universe with negligible gravity:
$$ds^2=\tau^2 dv^2-(dx^2+dy^2+dz^2).\eqno(3.1)$$
It is equal to zero for the Hubble expansion but is not vanishing at cosmic
times smaller than $\tau$. 

Notice the similarity to Minkowskian line element 
$$ds^2=c^2 dt^2-(dx^2+dy^2+dz^2)\eqno(3.2)$$
which vanishes for light propagation but is different from zero for 
particles of finite mass.\newline
5. Postulates of cosmological special relativity: (1) The laws of physics are
valid at all cosmic times; (2) $\tau$ has the same value at all cosmic times.
(Similarly to Einstein's special relativity postulates.)\newline
6. The cosmological transformation. This is the analog to the Lorentz 
transformation. It relates physical quantities at different cosmic times 
(similarly to the Lorentz transformation that relates quantities at different
velocities):
$$x'=\frac{x-\left(\tau-t\right)v}{\sqrt{\left(t/\tau\right)\left(2-t/\tau\right)}},
\hspace{7mm}
\tau v'=\frac{\tau v-x\left(\tau-t\right)/\tau}{\sqrt{\left(t/\tau\right)\left(2-t/\tau\right)}},\eqno(3.3)$$
$$y'=y,\hspace{5mm} z'=z,$$
$0\leq t\leq \tau$, $t=0$ at the Big Bang, $t=\tau$, now.\newline
7. Example: Denote the temperature of the Universe at a cosmic time $t$ by
$T$ and that at present by $T_0$ (=2.73K), we then have 
$$T=T_0/\sqrt{\left(t/\tau\right)\left(2-t/\tau\right)}.\eqno(3.4)$$
At $t/\tau=1/2$ we get $T=3.15$K. (This result assumes negligible gravity and
needs a correction by a factor of 13, and thus the temperature at $t/\tau=1/2$
is 41K.)\newline
8. The galaxy cone. This is the analog to the light cone in Einstein's special 
relativity. It represents the expansion of the Universe with negligible 
gravity. It is a four-dimensional cone. The coordinates are the velocity and 
the other three coordinates are the space coordinates. At the cone surface the
Hubble expansion is represented, whereas the inner part of the cone represents 
events at cosmic times less than $\tau$. The similarity to the light cone is
remarkable. 
\begin{figure}[t]
\centering
\includegraphics[height=10cm]{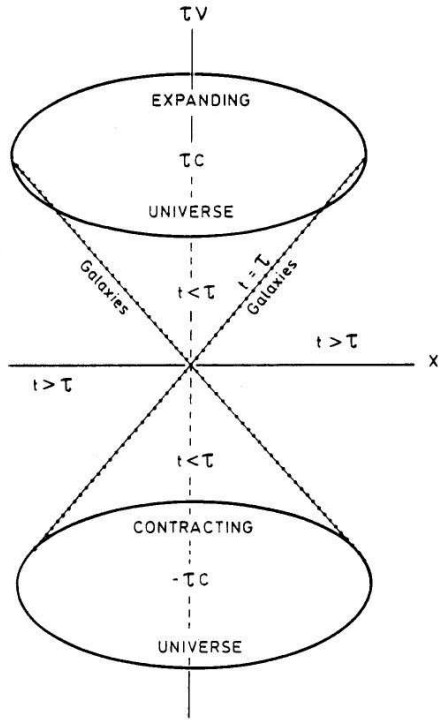}
\caption{The galaxy cone in cosmological relativity, describing the cone in
the $x-v$ space satisfying $\tau^2 v^2-x^2=0$, where $x$ represents the 
three-dimensional space. The heavy dots describe galaxies. The galaxy cone
represents the locations of the galaxies at a given time rather than their 
path of motion in the real space.}
\label{fig.3}
\end{figure}

9. Inflation at the early Universe. From the cosmological transformation we
obtain the relationship between the mass density $\rho_0$ now to its value
$\rho$ at a backward time $t$:
$$\rho=\rho_0/\sqrt{1-t^2/\tau^2}.\eqno(3.5)$$
The volume of the Universe is inversely proportional to its density, hence 
the ratio of the volumes at two backward cosmic times $t_1$ and $t_2$ is 
$$V_2/V_1=\sqrt{\left(1-t_2^2/\tau^2\right)/\left(1-t_1^2/\tau^2\right)}.\eqno(3.6)$$
For $t_1$, $t_2$ very close to $\tau$ we have
$$V_2/V_1=\sqrt{t'_2/t'_1},\eqno(3.7)$$
where primes indicate times with respect to the Big Bang. For $t'_2-t'_1
\approx 10^{-32}$s, and $t'_2$ much less than 1s, we have 
$$V_2/V_1=10^{-16}/\sqrt{t'_1}.\eqno(3.8)$$
For example, at $t'_1\approx 10^{-132}$s, we obtain $V_2/V_1\approx 10^{50}$
describing inflation.

The above introduction gives a brief review of a new special relativity
(cosmological special relativity, for more details see \cite{Carm02}). Obviously the
Universe is filled up with gravity and therefore one has to go to a 
Riemannian space with the Einstein gravitational field equations in terms of 
space and redshift (velocity). This is done in the second part of these 
lectures. Before that 
we outline Einstein's general relativity theory. 
\section{General relativity theory: A brief outline}
1. Postulates: (a) Principle of general covariance; (b) Principle of 
equivalence.\newline
2. Riemannian curved space for the gravitational field. The metric tensor 
$g_{\mu\nu}$ as the gravitational potential \cite{Carm82}.\newline
3. The Einstein field equations:
$$G_{\mu\nu}=R_{\mu\nu}-\frac{1}{2}g_{\mu\nu}R=\kappa T_{\mu\nu}\eqno(4.1)$$
4. The geodesic equation as the equation of motion:
$$\frac{d^2x^\rho}{ds^2}+\Gamma^\rho_{\alpha\beta}\frac{dx^\alpha}{ds}
\frac{dx^\beta}{ds}=0.\eqno(4.2)$$
\section{Extension to curved space}
The theory presented here, cosmological general relativity, uses a Riemannian 
four-dimensional presentation of 
gravitation in which the coordinates are those of Hubble, i.e. distances and 
velocity rather than the traditional space and time. We solve the field 
equations and show
that there are three possibilities for the Universe to expand. The theory 
describes the Universe as having a three-phase evolution with a decelerating
expansion, followed by a constant and  an accelerating expansion, and it 
predicts that the Universe is now in the latter phase. It is
shown, assuming $\Omega_m=0.245$, that the time at which the Universe goes over
from a decelerating to an accelerating expansion, i.e., the constant-expansion
phase, occurs at 8.5 Gyr ago. Also, at that time the cosmic radiation 
temperature was 146K. Recent
observations of distant supernovae imply that 
the Universe's growth is accelerating, contrary to what has always been 
assumed, that the expansion is slowing down due to gravity. Our theory 
confirms these recent experimental results by showing that the Universe now is 
definitely in a stage of accelerating expansion. The theory predicts also that 
now there is a positive pressure, $p=0.034g/cm^2$, in the Universe. It is
worthwhile mentioning that the theory has no cosmological constant. It is 
also shown that the three-dimensional space of the Universe is Euclidean, as 
the Boomerang, Maxima, DASI and CBI microwave telescopes have shown. 
Comparison with general relativity theory is
finally made and it is pointed out that the classical experiments as well as the 
gravitational radiation prediction follow from the present theory, too.
\section{Cosmology in spacevelocity}
In the 
framework of cosmological general relativity (CGR) gravitation is described
by a curved four-dimensional Riemannian spacevelocity. CGR incorporates the
BB constant $\tau$ at the outset. The Hubble law is assumed in CGR as a
fundamental law. CGR, in essence, extends Hubble's law so as to incorporate 
gravitation in it; it is actually a {\it distribution theory} that relates distances 
and velocities between galaxies. The theory involves measured quantities
and it takes a picture of the Universe as it is at any moment. The following 
is a brief review of CGR as was originally given by the author 
in 1996 \cite{Carm96c}.

The foundations of any gravitational theory are based on the principles of
equivalence and general covariance. These two principles lead immediately
to the realization that gravitation should be described by a four-dimensional
curved spacetime, in our theory spacevelocity, and that the field equations 
and the equations of motion should be generally covariant.
Hence these principles were adopted in CGR also. Use is made in a 
four-dimensional Riemannian manifold with a metric $g_{\mu\nu}$ and a line 
element $ds^2=g_{\mu\nu}dx^\mu dx^\nu$. The difference from Einstein's general
relativity is that our coordinates are: $x^0$ is a velocitylike coordinate 
(rather than a timelike coordinate), thus $x^0=\tau v$ where $\tau$ is the
Big Bang time and $v$ the velocity. The coordinate $x^0=\tau v$ is the 
comparable to $x^0=ct$ in
ordinary general relativity. The other three coordinates $x^k$, $k=1,2,3$, are
spacelike, just as in general relativity theory.

An immediate consequence of the above choice of coordinates is that the null
condition $ds=0$ describes the expansion of the Universe in the curved 
spacevelocity (generalized Hubble's law with gravitation) as compared to the 
propagation of light in the curved spacetime in general relativity. This means
one solves the field equations (to be given in the sequel) for the metric 
tensor, then from the null condition $ds=0$ one obtains immedialety the 
dependence of the relative distances between the galaxies on their relative
velocities.

As usual in gravitational theories, one equates geometry to physics. The first 
is expressed by means of the Einstein tensor. The physical part is expressed 
by the energy-momentum tensor which now has a different physical meaning from 
that in Einstein's theory. More important, the coupling constant that relates 
geometry to physics is now also {\it different}. 

Accordingly the field equations are
$$R_{\mu\nu}-\frac{1}{2}g_{\mu\nu}R=\kappa T_{\mu\nu},\eqno(6.1)$$
exactly as in Einstein's theory, with $\kappa$ given by
$\kappa=8\pi k/\tau^4$, (in general relativity it is given by $8\pi G/c^4$),
where $k$ is given by $k=G\tau^2/c^2$, with $G$ being Newton's gravitational
constant, and $\tau$ the Big Bang constant time. When the equations of motion 
are written in terms of velocity instead of time, the constant $k$ will
replace $G$. Using the above equations one then has $\kappa=8\pi G/c^2\tau^2$.

The energy-momentum tensor $T^{\mu\nu}$ is constructed, along the lines of
general relativity  theory, with the speed of light being replaced by
$\tau$. If $\rho$ is the average mass density of the Universe,
then it will be assumed that $T^{\mu\nu}=\rho u^\mu u^\nu,$ where $u^\mu=
dx^\mu/ds$ is the four-velocity.
In general relativity theory one takes $T_0^0=\rho$. In Newtonian gravity one
has the Poisson equation $\nabla^2\phi=4\pi G\rho$. At points where $\rho=0$
one solves the vacuum Einstein field equations in general relativity and the 
Laplace equation 
$\nabla^2\phi=0$ in Newtonian gravity. In both theories a null (zero) solution
is allowed as a trivial case. In cosmology, however, there exists no situation
at which $\rho$ can be zero because the Universe is filled with matter. In
order to be able to have zero on the right-hand side of (6.1) one takes 
$T_0^0$ not as equal to $\rho$, but to $\rho_{eff}=\rho-\rho_c$, where 
$\rho_c$ is the critical mass density, 
a {\it constant} in CGR given by $\rho_c=3/8\pi G\tau^2$, whose value is 
$\rho_c\approx 10^{-29}g/cm^3$, a few hydrogen atoms per cubic meter. 
Accordingly one takes
$T^{\mu\nu}=\rho_{eff}u^\mu u^\nu$; $\rho_{eff}=\rho-\rho_c$
for the energy-momentum tensor. Moreover, the above choice of the 
energy-momentum tensor is the only possibility that yields a constant
expansion when $\rho=\rho_c$ as it should be.

In Part II we apply CGR to obtain the accelerating expanding 
Universe and related subjects.

\end{document}